\documentclass[amsmath,amssymb,twocolumn,letterpaper]{revtex4}

\usepackage{QCMacros}
\ifx\pdftexversion\undefined
  \usepackage[dvips]{graphics}
\else
  \usepackage[pdftex]{graphicx}
\fi

\newcommand{\sing}{\ensuremath{\op{\rho}_+}}
\newcommand{\nrho}[1]{\ensuremath{\rho_{#1}}}
\newcommand{\nsig}{\hat{n}\cdot\vec{\sigma}}

\begin{document}
\title{Noise induced loss of entanglement}
\author{Kovid Goyal}%
  \email{physics@kovidgoyal.cjb.net}%
  \homepage{http://kovidgoyal.cjb.net}%
  \affiliation{St. Xavier's College, Mumbai 400001}


\begin{abstract}
  The disentangling effect of repeated applications of the bit flip
  channel ($\1\otimes\sigma_x$) on bipartite qubit systems is
  analyzed. It is found that the rate of loss of entanglement
  is not uniform over all states. The distillable entanglement of
  maximally entangled states decreases faster than that of less
  entangled states. The analysis is also generalized to noise channels
  of the form $\nsig$. 
\end{abstract}

\maketitle
\allowdisplaybreaks 

\section{Introduction}
\begin{figure}[ht]
  \centering
  \begin{picture}(100,80)
    \put(10, 10){\vector(1, 0){80}}
    \dtext{2}{6}{1}\dtext{92}{6}{1}
    \dtext{2}{66}{0}\dtext{92}{66}{0}
    \put(10, 70){\vector(1, 0){80}}
    \dtext{37}{55}{$1-p$}\dtext{37}{20}{$1-p$}
    \dtext{46}{75}{$p$}\dtext{46}{3}{$p$}
    \put(20, 10){\vector(1, 1){60}}
    \put(20, 70){\vector(1, -1){60}}
  \end{picture}
  \caption{The binary symmetric channel}
  \label{fig:bin-symm-chann}
\end{figure}
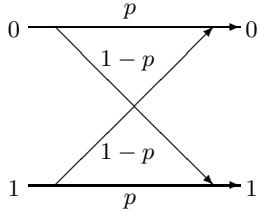

The storage/transmission of classical data is subject to various noise
processes that reduce the integrity of the data over time. One such
noise process is the \textit{binary symmetric channel}
(\Fref{fig:bin-symm-chann}), that flips a bit with a given probability
$1-p$. There exist many, successful strategies for dealing with this noise process
\cite{Welsh:1988}. 

\textit{Entanglement} is a quantum resource, essential to many
applications such as teleportation, super-dense coding, etc. As such,
the ability to combat noise during the storage of entanglement is
essential. In this paper, we consider an instance of the binary
symmetric channel, applied to bipartite qubit systems. We analyze the
disentangling effect of this channel on singlet (maximally entangled)
states. The choice of a qubit system is dictated by the existence of a
mathematically tractable measure of entanglement for bipartite qubit
systems \cite{Wootters:1998}.

\section{The Quantum Bit Flip Channel}

The generalization of the symmetric bit flip channel to the case of a
single qubit is straightforward. Choose the computational basis
($\{\ket{0},\ket{1}\}$) of the Hilbert space $\op{H}_2$. Let $\rho$ be
any density matrix acting on this space. Then the quantum bit flip
channel can be defined as
\begin{equation}
  \label{eq:sq-bf-defn}
  \rho' = p\rho + (1-p)\sigma_x\rho\sigma_x.
\end{equation}
In order to study the effect of this channel on entanglement, this
definition needs to be extended for bipartite systems. We make the
choice that only one of the two subsystems is affected by the
noise. Then for $\rho \in \op{H}_2\otimes\op{H}_2$,
\begin{align}
  \label{eq:bipart-bf-defn}\notag
  \rho' &= p\rho + (1-p)\op{X}(\rho),\\
  \op{X}(\rho) :&= (\1\otimes\sigma_x)\rho(\1\otimes\sigma_x).
\end{align}
Since $\sigma_x$ is a completely positive map, $\rho'$ is also a
density matrix in $\op{H}_2\otimes\op{H}_2$. We are interested in the
disentangling effect of this channel on the maximally entangled
singlet state, defined as
\begin{equation}
  \label{eq:sing-defn}
  \sing = \frac{1}{2}\sum_{i,j=0}^1 \ketbra{i}{j}\otimes\ketbra{i}{j}.
\end{equation}

After a single application of the channel, the resulting density
matrix \nrho{1} has the form
\begin{equation}
  \label{eq:rho1-defn}
  \nrho{1} = p\sing + (1-p)\op{X}(\sing).
\end{equation}
The entanglement of this state should be a function of $p$, which
completely parameterizes the bit flip channel.
\subsection{Entanglement of Formation}
\begin{figure}[ht]
    \includegraphics{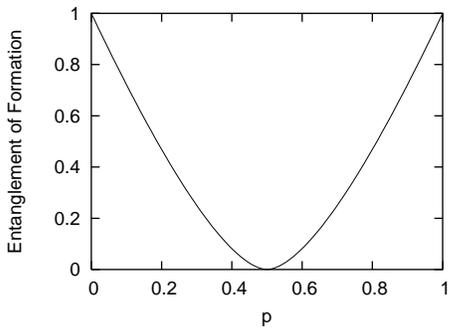}
  \caption{Entanglement of $\rho_{(1)}$ as a function of $p$}
  \label{fig:rho-one-stage}
\end{figure}

In order to calculate the
entanglement of formation \cite{Wootters:1998}, the following
definitions are required
\begin{align}
  \label{eq:ent-defs}
  \tilde{\rho}  &= (\sigma_y\otimes\sigma_y) \rho^* (\sigma_y\otimes\sigma_y), \\
  \op{C}(\rho)  &= \max \{0, \lambda_1 - \lambda_2 - \lambda_3 - \lambda_4\},
\end{align}
where $\lambda_1 \geq \lambda_2 \geq \lambda_3 \geq \lambda_4$ are the
the eigenvalues of the matrix
$\sqrt{\sqrt{\rho}\tilde{\rho}\sqrt{\rho}}$. Then the entanglement of
formation $\E(\rho)$ is given by
\begin{align}\notag
  \label{eq:eof-defn}
  \E(\rho) &= h\left(\frac{1+\sqrt{1-\op{C}(\rho)^2}}{2}\right),\\
  h(x) &= -x\log_2x - (1-x)\log_2(1-x).
\end{align}
For \nrho{1} the concurrence is found to be
\begin{equation}
  \label{eq:rho1-con}
  \op{C}(\nrho{1}) = 2\left|p-\frac{1}{2}\right|.
\end{equation}
This gives an entanglement of formation
\begin{equation}
  \label{eq:form-defn}
  \E_F(\nrho{1}) = h\left(\frac{1}{2} + \sqrt{p(1-p)}\right).
\end{equation}
An outline of the calculations is presented in
\Sref{sec:gen}. \Fref{fig:rho-one-stage} shows how the entanglement
varies as a function of $p$.

\subsection{Distillable Entanglement}
While there doesn't exist a general method for calculating the
distillable entanglement of an arbitrary density matrix, we are
fortunate in that \nrho{1} can be written in the Bell diagonal form,
as
\begin{equation}
  \label{eq:rho1-bell-diag}
  \nrho{1} = p\ketbra{\Phi^+}{\Phi^+} + (1-p)\ketbra{\Psi^+}{\Psi^+}.
\end{equation}
Using the one way hashing protocol \cite{BDSW:1996} for distillation, it is possible to obtain a
lower limit on the distillable entanglement of $1-h(p)$.
The distillable entanglement is bound above by the relative
entropy of entanglement \cite{Vedral/Plenio:1998}, which for \nrho{1}
is also \cite{VPRK:1997}, $1-h(p)$.
Combining the two bounds, we have
\begin{equation}
  \label{eq:dist-defn}
  \E_D(\nrho{1}) = 1 - h(p).
\end{equation}

\subsection{Multiple Applications}
\begin{figure}[ht]
  \includegraphics{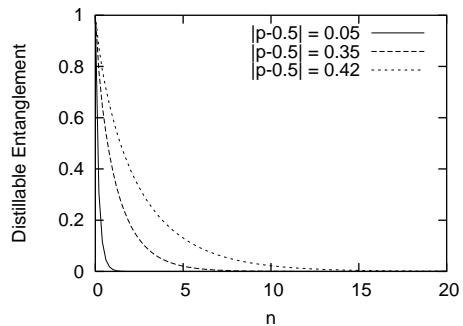}
  \caption{Entanglement of $\rho_{(n)}$ for $|p-\frac{1}{2}| = 0.05,
    0.35, 0.42$. The curves have been smoothed by calculating
    \Eref{eq:ents-fin} for non integral values of $n\in[0,20]$.}
  \label{fig:eof-n}
\end{figure}

We now ask the question, what effect do multiple applications of the
channel have on the singlet state? In order to answer it,
we need to know the form of the singlet state after $n$ applications,
denoted by \nrho{n}. Proceeding from \Eref{eq:rho1-defn},
\begin{align}\notag
  \nrho{2} &= p\nrho{1} +(1-p)\op{X}(\nrho{1})\\\label{eq:rho2-form}
  &= (p^2 + (1-p)^2)\sing + (p(1-p) + (1-p)p)\op{X}(\sing)\\\notag
  &= P_2\sing + (1-P_2)\op{X}(\sing);\\\notag
  P_2 &= p^2 + (1-p)^2.
\end{align}
The identity $\sigma_x^2=\1$ was used to arrive at
\Eref{eq:rho2-form}. Thus \nrho{2} has exactly the same form as
\nrho{1}; repeated applications of the channel will not change this
form. All that remains is to find an expression for $P_n$. By
calculating \nrho{n} for the first few $n$ explicitly, we have
\begin{align*}
  &\begin{split}
      P_0 &= 1,  \\ P_1 &= p,
    \end{split}
    &\begin{split}
      P_2 &= p^2 + (1-p)^2, \\ P_3 &= p^3 + 3p(1-p)^2.
    \end{split}
\end{align*}
Evidently, $P_n$ is the sum of the even terms from the expansion of
$(p+(1-p))^n$. 
\begin{align}\notag
  \therefore P_n &= \frac{(p+(1-p))^n + (p-(1-p))^n}{2}\\\label{eq:P_n-defn}
  &= \frac{1}{2} + 2^{n-1}\left(p-\frac{1}{2}\right)^n.
\end{align}
Now that we have obtained a general expression for $P_n$, we can
calculate the entanglements as,
\begin{align}
  \label{eq:ents-fin}\notag
  \E_F(\nrho{n}) &= h\left(\frac{1}{2}+\sqrt{P_n(1-P_n)}\right)\\
  \E_D(\nrho{n}) &= 1-h(P_n).
\end{align}
\Fref{fig:eof-n} shows how the distillable entanglement
decreases with $n$ for different values of $\left|p-\frac{1}{2}\right|$.

\subsection{Combating the Disentanglement}
The form of the curves in \Fref{fig:eof-n} suggests that perhaps,
states further along the curves lose entanglement slower than the
singlet. In order to test this, first we define the fractional loss of
entanglement the state \nrho{k} after $r$ applications of the channel as
\begin{equation}
  \label{eq:frac-defn}
  F(p, k, r) = - \frac{\E(\nrho{k}) - \E(\nrho{k+r})}{\E(\nrho{k})};
\end{equation}
where $\E(\rho)$ is a measure of the entanglement of $\rho$. Then the
fractional loss of entanglement of the singlet state after $r$
applications of the channel is given by $F(p, 0, r)$. In order to
compare the loss of entanglement of the singlet state with that of
\nrho{k}, define
\begin{equation}
  \label{eq:frac-adv-defn}
  R(p, k, r) = \frac{F(p, k, r)}{F(p, 0, r)}.
\end{equation}
\begin{center}
  \begin{widetext}
    \hspace{1.9cm}
    \begin{figure}[ht]
      \centering
      \begin{tabular}{cc}
        \includegraphics{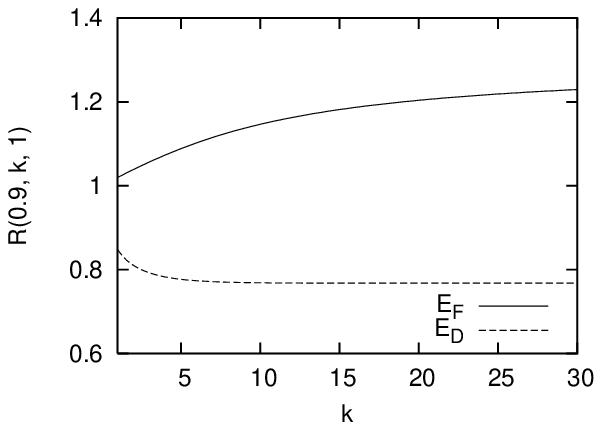} & 
        \includegraphics{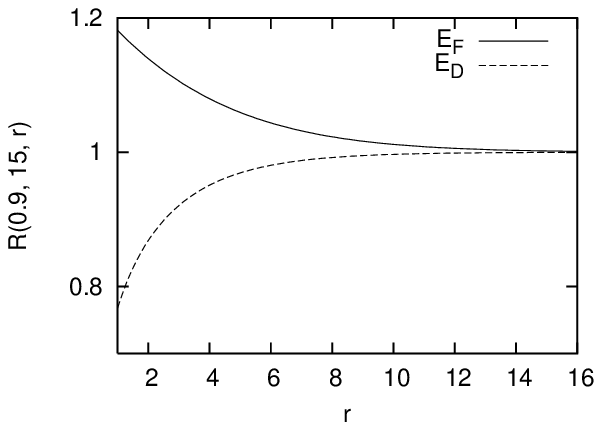}\\
        \multicolumn{2}{c}{\includegraphics{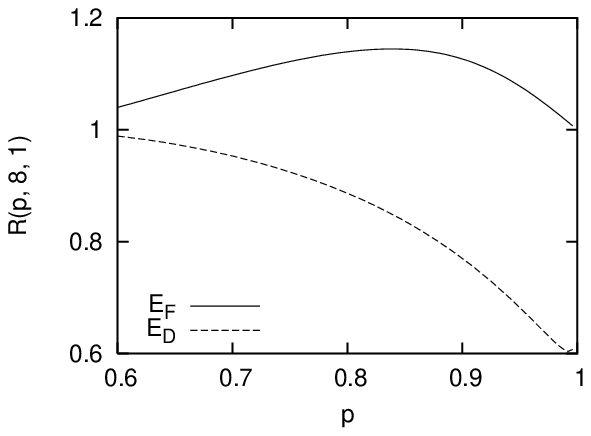}}
      \end{tabular}      
      \caption{Graphs showing the dependence of $R(p, k, r)$ on $p, k$ and
        $r$. It is seen that $R$ behaves differently for different
        measures of entanglement.}
      \label{fig:disen-stuff}
    \end{figure}
  \end{widetext}
\end{center}
\Fref{fig:disen-stuff} illustrates the behavior of $R(p,k,r)$. The
most striking feature of the graphs is that the entanglement of
formation and the distillable entanglement behave in a qualitatively
different manner with regard to the rate of loss of entanglement of
\nrho{k}.The rate of loss of entanglement of formation is higher for
\nrho{k} than for the singlet state. The reverse is true for the
distillable entanglement. 

It is the distillable entanglement that is of greater practical
interest, and the fact that \nrho{k} loses it slower than the singlet
suggests a simple tactic to combat the disentangling action
of this channel. Rather than storing entanglement as a few
singlets, it should be stored as a larger number of less
entangled states of the form of \nrho{k}. Since the fractional loss of
entanglement for these states is less than for the singlet, there will
be a smaller net loss of entanglement over time, \textit{provided}
that the distillable entanglement for these states is additive, that
is
\begin{equation}
  \label{eq:additive}
  \E_D(\nrho{k}^{\otimes N}) = N\E_D(\nrho{k}).
\end{equation}
This will ensure that the entanglement spread over $N$ copies of these
states can be efficiently concentrated into singlet form again.

The second graph in \Fref{fig:disen-stuff} shows that the advantage
obtained by storing the entanglement in dilute form is lost if the
system is exposed to noise repeatedly. While this does impose a limit
on the savings that can be made, if a sufficiently large $k$ is
chosen and $r$ is bounded, there can still be significant gains.

The final graph shows, rather predictably, that  the less severe the
noise, the greater the gains that can be made, for a given $k$ and $r$.

\section{Generalization}
\label{sec:gen}
Although most of the results in this paper are derived for
the bit flip channel, a number of them hold for more general noise
processes as well. In this section, we will analyze the general noise
process 
\begin{align}
  \label{eq:rho1-gen-defn}\notag
  \nrho{1} &= p\sing + (1-p)\op{N}(\sing)\\\notag
  \op{N}(\rho) :&= (\1\otimes\nsig)\rho(\1\otimes\nsig)\\
  & = \frac{1}{2}\sum_{i,j=0}^1\sum_{a,b=1}^3n_an_b\ket{i}\bra{j}\otimes\sigma_a \ket{i}\bra{j}\sigma_b.
\end{align}
where $\hat{n}\in\mathbb{R}^3$ is arbitrary. For $\hat{n} = (1,0,0),
(0,1,0)$ and $(0,0,1)$, this channel reduces to the bit flip,
bit-phase flip and phase flip channels respectively \cite{Nielsen/Chuang:2000}.

\subsection{Entanglement of Formation}
Here we explicitly calculate the entanglement of formation of
\nrho{1}, defined in \Eref{eq:rho1-gen-defn}. First we need to
evaluate $\tilde{\nrho{1}} =
(\sigma_y\otimes\sigma_y)\nrho{1}(\sigma_y\otimes\sigma_y)$.
The following identity \cite{Jozsa:1994}, comes in handy
\begin{equation}
  \label{eq:jozsa-ident}
  (\1\otimes M)\sing(\1\otimes M^\dagger) = (M^T\otimes\1)\sing(M^*\otimes\1);
\end{equation}
where $M$ is any matrix. As a result of \Eref{eq:jozsa-ident} we get
\begin{align}\label{eq:inv1}\notag
  (\sigma_y&\otimes\sigma_y)\sing^*(\sigma_y\otimes\sigma_y)\\\notag
  &= (\sigma_y\otimes\1)(\1\otimes\sigma_y)\sing^*(\1\otimes\sigma_y)(\sigma_y\otimes\1)\\
  &= \sing^* = \sing.
\end{align}
Define $\hat{n}' = (n_x, -n_y, n_z)$. Then, for the second term in \nrho{1}
\begin{align}
  \label{eq:inv2}\notag
  (\sigma_y&\otimes\sigma_y)\op{N}(\sing)^*(\sigma_y\otimes\sigma_y)\\\notag
  &=(\sigma_y\otimes\sigma_y)(\1\otimes\hat{n}'\cdot\vec{\sigma})
  \sing(1\otimes\hat{n}'\cdot\vec{\sigma})(\sigma_y\otimes\sigma_y)\\\notag
  &=\frac{1}{2}\sum_{a,b,i,j}\sigma_y\ket{a}\bra{b}\sigma_y\otimes
  n_i'\sigma_y\sigma_i\ket{a}\bra{b}n_j'\sigma_j\sigma_y\\\notag
  &=\frac{1}{2}\sum_{a,b,i,j}\sigma_y\ket{a}\bra{b}\sigma_y\otimes
  (-n_i)\sigma_i\sigma_y\ket{a}\bra{b}(-n_j)\sigma_y\sigma_j\\\notag
  &=(\1\otimes\nsig)(\sigma_y\otimes\sigma_y)\sing(\sigma_y\otimes\sigma_y)(\1\otimes\nsig)\\
  &=\op{N}(\sing).
\end{align}
\Eref{eq:inv1} and \Eref{eq:inv2} together imply that
$\tilde{\nrho{1}} = \nrho{1}$. Thus in order to calculate the
concurrence of \nrho{1} we need to know only its eigenvalues. The
matrix is
\begin{widetext}
  \begin{align}
    \label{eq:gen-rho1-defn}
    &\nrho{1} = \frac{1-p}{2}\left[
      \begin{matrix}
        r + n_z^2     & (n_x-in_y)n_z  & (n_x + in_y)n_z & r-n_z^2 \\
        (n_x+in_y)n_z & n_x^2 + n_y^2  & (n_x + in_y)^2  & -(n_x+in_y)n_z\\
        (n_x-in_y)n_z & (n_x-in_y)^2   & n_x^2 + n_y^2   & -(n_x-in_y)n_z \\
        r-n_z^2       & -(n_x-in_y)n_z & -(n_x+in_y)n_z  & r+n_z^2
      \end{matrix}\right];
    &r  = \frac{p}{1-p}.
  \end{align}
\end{widetext}
Amazingly enough, the eigenvalues of this matrix are $\{p, 1-p,0,0\}$
giving a concurrence
\begin{equation}
  \label{eq:gen-con}
  \op{C} = |2p-1|.
\end{equation}
This is the same result as was obtained for the bit flip channel in
\Eref{eq:rho1-con}. The fact that $(\nsig)^2 = \1$ ensures that
\begin{equation}
  \label{eq:gen-n-form}
  \nrho{n} = P_n\sing + (1-P_n)\op{N}(\sing).
\end{equation}
It can easily be demonstrated that this $P_n$ is the same as was obtained
for the bit flip channel in \Eref{eq:P_n-defn}. Thus, the analysis
carries over entirely for the $\nsig$ channel, in the case of
entanglement of formation. 

For the distillable entanglement, the situation is complicated by the
absence of any method for calculating the entanglement for an
arbitrary density matrix. However, for the special cases of
$\hat{n}=(1,0,0), (0,1,0)$ or $(0,0,1)$ \nrho{1} remains in Bell
diagonal form. As a result its distillable entanglement is easily
calculated to be $1-h(p)$, as in \Eref{eq:dist-defn}.

\section{Conclusion}
Noise reduces bipartite entanglement (of a singlet) exponentially, at a rate that
depends on how non uniform the noise probability is. The greater the
distance of the noise probability $p$ from $1/2$, the less severe the
noise. While the noise never totally destroys the entanglement, it
does make it negligible very quickly. 

Interestingly, noise seems to affect states differently. The
distillable entanglement of the singlet reduces faster than that of
\nrho{k}. Theoretically, this is interesting behavior in
itself. There seems to be no \textit{a priori} reason why the singlet
should be more fragile than its less entangled counterparts.
Practically, it is of importance as it suggests that
entanglement should not be stored in the form of singlets.

The rate of loss of entanglement of formation was found to be the same
for the generalized $\nsig$ channel as that for the
$\1\otimes\sigma_x$ channel. The rate of loss of distillable
entanglement for the special cases $\hat{n}=(1,0,0),(0,1,0)$ and
$(0,0,1)$ was uniformly $1-h(p)$. It is conjectured that this is the
rate of loss of distillable entanglement for arbitrary $\hat{n}$.

\begin{acknowledgments}
  I would like to thank Dr. R. Simon for his advice and for many
  stimulating discussions. I would also like to thank Dr. Ajay
  Patwardhan for his support and guidance over the years. I also
  acknowledge the support in the form of a Summer Fellowship from the
  Indian Academy of Sciences and Institutional support from the
  Institute of Mathematical Sciences, Chennai, without which this
  paper would never have been written.
\end{acknowledgments}

\bibliography{QuantumComputing}

\end{document}